\documentclass[aps,pre,floatfix,twocolumn]{revtex4}
\usepackage{graphicx}

\usepackage{amsfonts,amsmath,amssymb,amsthm}
\def\beq{\begin{equation}}
\def\eeq{\end{equation}}
\def\bea{\begin{eqnarray}}
\def\eea{\end{eqnarray}}

\begin{document}
\makeatletter
\title{Provoking Predetermined Aperiodic Patterns in Human Brainwaves}

\author{Richa Phogat}
\affiliation{Department of Physics,
Indian Institute of Technology,
Bombay, Powai, Mumbai--400 076, India.}

\author{P. Parmananda}
\affiliation{Department of Physics,
Indian Institute of Technology,
Bombay, Powai, Mumbai--400 076, India.}

\begin{abstract}

In the present work, electroencephalographic recordings of healthy human participants were performed to study the entrainment of brainwaves using a variety of stimulus. First, periodic entrainment of the brainwaves was studied using two different stimuli in the form of periodic auditory and visual signals. The entrainment with the periodic visual stimulation was consistently observed, whereas the auditory entrainment was inconclusive. Hence, a photic (Visual) stimulus, where two frequencies were presented to the subject simultaneously was used to further explore the bifrequency entrainment of human brainwaves. Subsequently, the evolution of brainwaves as a result of an aperiodic stimulation was explored, wherein an entrainment to the predetermined aperiodic pattern was observed. These results suggest that aperiodic entrainment could be used as a tool for guided modification of brainwaves. This could find possible applications in processes such as epilepsy suppression and biofeedback. 
\end{abstract}

\maketitle

{\bf Variations in the environment of a system can alter its dynamics. In this work, the phenomenon of the entrainment of human brainwaves to a predetermined aperiodic photic stimulus is presented. As a precursor, entrainment to a single frequency and a bi-frequency signal is studied. Subsequently, the presence of aperiodic entrainment was quantified using short-time Fourier transform (STFT) and by calculating the cross correlation between the STFT of the aperiodic signal and that of the observed EEG dynamics. This guided modification of brainwaves may find possible applications in suppressing some types of epilepsy and in biofeedback.}

Entrainment is the process of adjusting the dynamics of a system to that of an external rhythm. This is observed in a wide variety of natural as well as laboratory systems~\cite{Parmananda, Zlotnik, Klein, Freedman, Stokkan, Takahashi}. In mammals, entrainment of the circadian rhythms as a function of various factors such as illumination, body temperature, social cues and food availability is well documented in literature~\cite{Klein, Freedman, Stokkan, Takahashi}. Another interesting observation in this field is the phenomenon of brainwave entrainment~\cite{Notbohm}. This phenomenon leads back to the initial experiments done to study the brain dynamics~\cite{Adrian, Walter}, wherein flickering lights at different frequencies were used to study the modification of brainwaves in human as well as animal subjects. A recent interest has emerged in the entrainment of brainwaves using a variety of stimulation such as, audio-visual stimulation (AVS) or transcranial alternating current stimulation and its possible applications~\cite{Aftanas, Helfrich, Teplan2, Teplan3}. Research has also been carried out to study the individual effects of auditory~\cite{Karino, Neher, Will} and photic~\cite{Mori} entrainment of the brainwaves. Noise along with a subthreshold photic stimulus has previously been shown to enhance the periodicity in brainwaves via stochastic resonance~\cite{Mori, Benzi}. However, there have been contradicting reports regarding the effects of auditory stimulation on the brainwave entrainment~\cite{Karino, Neher, Lopez}. The auditory stimulation is conventionally given in the form of binaural beats~\cite{Will, Teplan2, Teplan3, Karino, Lopez} or repeating drum sounds~\cite{Neher}. Similarly, the photic entrainment is studied using both colored~\cite{Teplan2, Teplan3} and white light~\cite{Mori} LEDs flickering at a desired frequency. In the present work, entrainment is studied using white light LEDs to avoid the psychological effects of the coloured light~\cite{Elliot}, if any. Also, the auditory counterpart of white light i.e. white sound was switched on and off periodically to explore the auditory entrainment of brainwaves.\\
The next step after studying single frequency entrainment would be bifrequency entrainment, wherein two rhythmic photic stimuli are presented to the subject simultaneously. Recently, in synthetic gene oscillators, entrainment was studied using aperiodic signals ~\cite{Butzin}. Another interesting extension to be explored in this direction would be an aperiodic entrainment of human brainwaves. We have studied the effects of an aperiodic photic stimulation on the electrical activity of the brain.
The organization of this paper is as follows. In part I, the protocol employed for the experiments is described. In part II, results for the periodic photic stimulation are presented. Bifrequency and aperiodic entrainment are reported in part III and IV respectively. A discussion on the results follows in part V.

\section{Experimental Protocol}
The experiments were performed on five healthy adults (26.12$\pm$1.86) who volunteered for the experiments. All participants were informed about the experimental protocol beforehand and the experiments were performed only after the participants signed the Informed Consent Form(ICF). The ICF was approved by the institute ethics committee of IIT Bombay. A set of 20 cerebral electrodes with the help of a adhesive electrode paste were used to record the EEG data at a sampling frequency of 256 Hz. The 10-20 electrode placement system was used for the positioning of the electrodes on the scalp~\cite{Homan}. Four additional electrodes were used to capture the eye movement and heart beat. Fpz was used as the reference and nasion was grounded. The data recorded was first cleaned visually for the artefacts in the EEGlab~\cite{Delorme} toolbox of MATLAB\textsuperscript{\textregistered} and then analysed using in house MATLAB\textsuperscript{\textregistered} codes.\\

The photic stimulus used for the experiments is a square wave of required frequency distribution, created using MATLAB\textsuperscript{\textregistered}. This digital square signal was supplied using a digital to analogue converter (MCC USB-1616HS-4) to a set of 8 LEDs mounted on a board. When maintained in a state of constant illumination, without any flicker, these LEDs have an intensity of 1125$\pm$98 Lux at a distance of $\approx$ 78 cm from the LED board. This was the distance maintained between the LEDs and the eyes of the subject. Therefore, when the stimulation is provided, the LEDs flicker with the given frequency distribution between the maximum illumination of 1125$\pm$98 Lux and a minimum illumination of 0 Lux. The subjects were sitting comfortably in a chair with their eyes closed to ensure alpha range (8-13 Hz) as the baseline~\cite{Adrian, Dolce}. The experiments were performed in a dark space so that the photic stimulus provided was the only source of light. The protocol for all the experiments was as follows:

0-8 minutes : Relaxed state (Part I)\\
8-18 minutes : Stimulus applied (Part II)\\
18-26 minutes : Relaxed state (Part III)\\
26-36 minutes : Stimulus applied (Part IV)\\
36-44 minutes : Relaxed state (Part V)

\section{Periodic Photic Stimulation}
The light signal was provided using a set of 8 white light LEDs mounted on a board. The first set of experiments were performed with all these LEDs synchronously flickering at 10 Hz frequency. This value of frequency was chosen to check for the periodic entrainment of the brainwaves with a stimulation in the baseline frequency range (alpha range: 8-13 Hz). Scalp maps were used to ensure maximum entrainment in the occipital head region. The analysis of the results was then performed using the short-time Fourier transform (STFT) to see the evolution of brainwaves in both time as well as the frequency domain. The spectrogram function of MATLAB\textsuperscript{\textregistered} was used for this purpose. The STFT was calculated using a Gaussian window of 6 s with an overlap of 5.96 s between consecutive windows. The EEG recordings for the Oz electrode were used for the analysis. This was done considering the symmetric location of this electrode between the left and right hemispheres, thus minimizing the effects of right or left handedness of the subject, if any. To quantify the entrainment observed, $\zeta$ is calculated and compared across various experimental conditions for Oz electrode. It gives a measure of the increment in the power of the Oz electrode when the photic stimulus is provided as compared to when no stimulus is given.

$\zeta = \int_{f_1}^{f_2} \int_{t_1}^{t_2} P(f,t) df dt$

$P$ represents the power spectral density (PSD) in frequency $f$ at time $t$. Unless otherwise specified, the values of $f_1$ and $f_2$ were kept constant at 5 Hz and 40 Hz respectively. This was done to study only the fundamental and super-harmonic entrainment of brainwaves. This frequency range includes all the discernible changes at fundamental and harmonic frequency observed with the entrainment. First two minutes of the cleaned EEG data were used for the analysis across the subjects and for all the experiments. Hence, $t_1$ = 0 s and $t_2$ = 2 min. To study the effects of the stimulation on the amplitude of the brainwaves, the variance in the amplitude with and without stimulus was compared. The amplitude variance was calculated as follows:

$Amplitude Variance=\frac{1}{N-1} \sum_{i=1}^{N} |A_i - \mu|^2$,

Where N is the total number of data points in the time-series A and $\mu = \frac{1}{N} \sum_{i=1}^{N} A_i$ is the mean of A.\\
The effect of the 10 Hz photic stimulation on the brainwaves calculated using the techniques mentioned above are presented in Figure 1 and Figure 2. Figure 1 shows the STFT for one of the subjects without (upper panel) and with (lower panel) the stimulus. An increment in the power across the fundamental frequency and the subsequent harmonics can be observed when the stimulus was provided as compared to when no stimulus was given. A discernible increment in $\zeta$ (left panel) and the amplitude variance (right panel) of the brainwaves when the stimulus is provided (Stimulus Status: $\bf{On}$) can be seen from Figure 2. Sub-harmonic entrainment of the brainwaves at this frequency was also studied by filtering the data in the frequency range of 4.9-5.1 Hz (sub-harmonic for 10 Hz is at 5 Hz) and studying both $\zeta$ and amplitude variance as a function of stimulus status. A persistent increment in both the quantities when the stimulus is provided was observed across all the subjects (results in supplementary material).\\

\begin{figure}[h!]
\includegraphics[width=9cm]{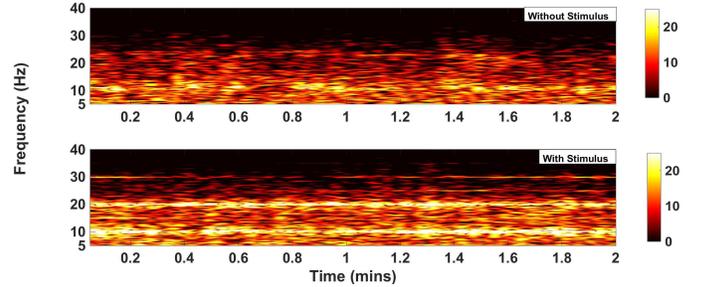}
\caption{STFT of the brainwaves (Oz electrode) for the first 2 minutes of without and with 10 Hz photic stimulation. An increment in power at 10 Hz and the subsequent harmonics (20 and 30 Hz) can be observed in the sub-plot where the stimulus is provided (lower panel).}
\label{1}
\end{figure}

\begin{figure}[h!]
\includegraphics[width=9cm]{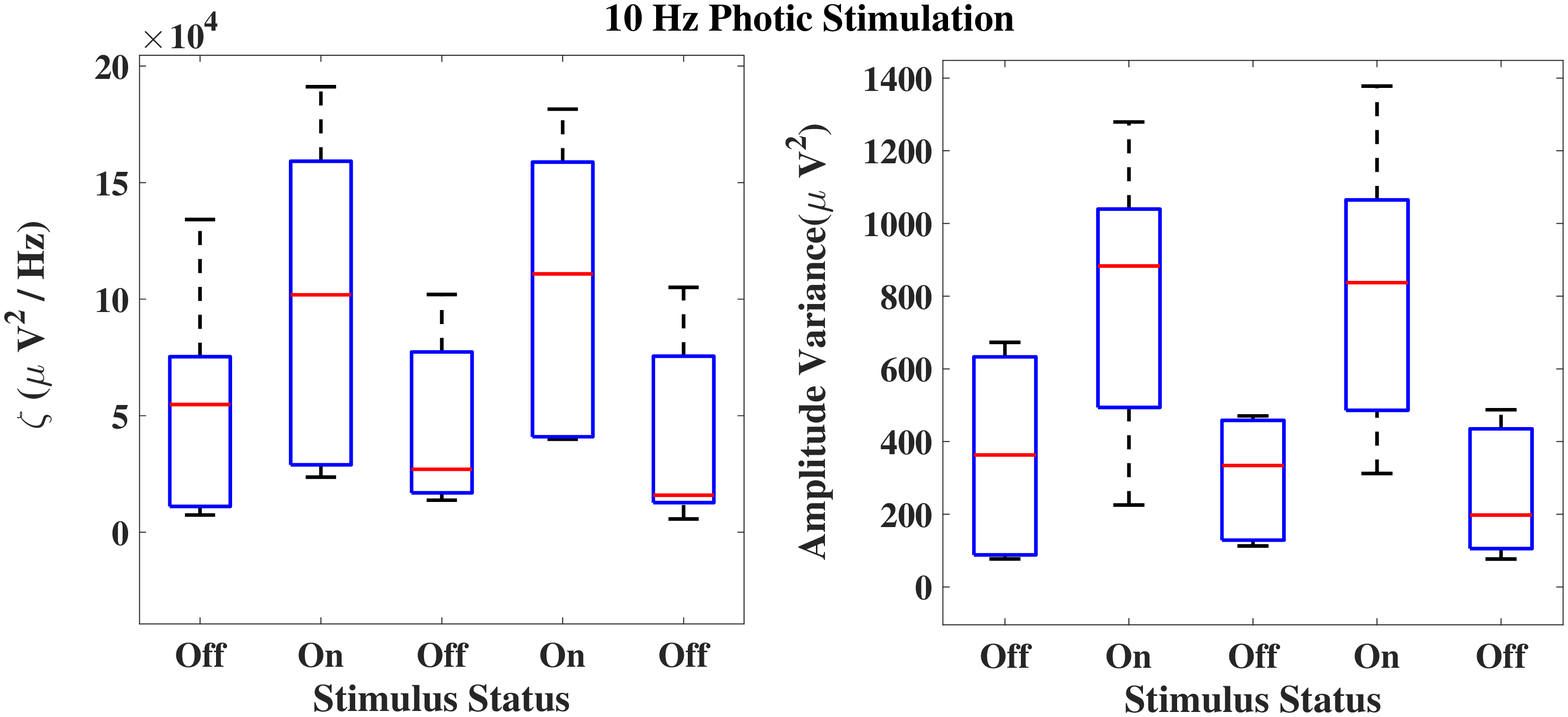}
\caption{Evolution of the quantities $\zeta$ and Amplitude Variance as a function of stimulus status for all the subjects. A universal increment for both these quantities when the stimulus is provided (Stimulus Status: $\bf{On}$) is evident from the box plots in both the sub-figures.}
\label{2}
\end{figure}

Upon observing entrainment using a 10 Hz photic stimulation, the effects of a 6 Hz photic stimulation on the brainwaves were explored. This was done to study brainwave entrainment when the entrainment frequency is relatively farther from the baseline frequency range. As shown in Figure 3, entrainment at fundamental frequency and its subsequent harmonics can be observed when the stimulus is provided (lower panel). However, the entrainment observed at 6 Hz is weaker as compared to the entrainment observed at 10 Hz. This is evident by lesser percentage increment in $\zeta$ and amplitude variance in Figure 4 as compared to Figure 2. Also, the sub-harmonic entrainment for 6 Hz calculated by filtering the brainwaves in the frequency range of 2.9 - 3.1 Hz (sub-harmonic for 6 Hz at 3 Hz), was not observed across all the subjects (results in supplementary material).\\

\begin{figure}[h!]
\includegraphics[width=9cm]{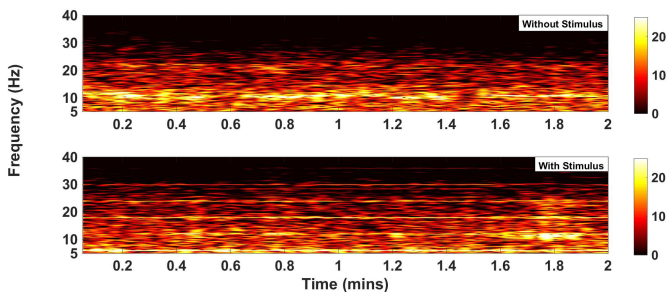}
\caption{STFT of the brainwaves (Oz electrode) for the first 2 minutes of without and with 6 Hz photic stimulation. An increment in power at 6 Hz and the subsequent harmonics (12, 18, 24 , 30 and 36 Hz) is detected in the sub-plot with the stimulus.}
\label{3}
\end{figure}

\begin{figure}[h!]
\includegraphics[width=9cm]{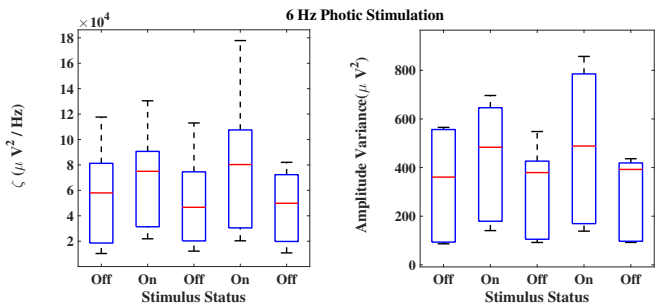}
\caption{Evolution of the quantities $\zeta$ and Amplitude Variance as a function of stimulus status for 6 Hz photic stimulus across all the subjects. A consistent increment in both these quantities when the stimulus is provided (Stimulus Status: $\bf{On}$) is indicated by the box plots in both the figures.}
\label{4}
\end{figure}

After studying photic entrainment, the next step was to study the effects of auditory stimulation on the brainwaves. For this purpose, an auditory analogue of white light i.e. white sound was used. This sound consists of the frequencies in the auditory range i.e. 22-22000 Hz distributed uniformly. The required auditory stimulation signal was then consturcted by oscillating between no sound and white sound at the desired frequencies (6 and 10 Hz). However, a consistent increment across all the subjects was not observed using this form of auditory stimulation. Amongst the subjects that did show entrainment, a significantly smaller increment in power was observed as compared to the photic stimulation (a detailed comparison for the same is presented in supplementary material). The results for auditory stimulation in this case are in agreement with those reported in~\cite{Lopez} for binaural beat entrainment.

\section{Bifrequency Entrainment}
Since a single frequency stimulus was able to entrain the brainwaves, two frequencies were provided simultaneously and the evolution of brainwaves was analysed. For this purpose, half the LEDs were flickering at 6 Hz and the other half at 10 Hz. An entrainment to both the frequencies as well as their subsequent harmonics as shown in Figure 5 was observed. Also, an entrainment to the sum of the two different frequencies (6+10 = 16 Hz) was observed. Figure 6 shows the consistency of this pattern across all the subjects for both $\zeta$ and amplitude variance. Bifrequency entrainment was further explored by supplying one frequency to each eye. A persistence of entrainment at the harmonics and summation of the two frequencies was observed. This indicates to the integration of the information received by each eye in the visual pathways of the brain. Since sub-harmonic entrainment was observed using 10 Hz photic stimulation and not with 6 Hz photic stimulation, it was not pursued for the analysis in the subsequent set of experiments. As seen in Figure 5, the bifrequency photic stimulation was able to simultaneously provoke a range of frequencies in the brainwaves. Subsequently, the time required to entrain the brainwaves was investigated. In~\cite{Notbohm}, it was observed that only a periodic stimulus provoked a higher strength of entrainment demonstrated by a better phase locking between the forcing and the brainwaves. This motivated us to find the minimum duration of periodic signal required for entrainment. This was checked by alternating the forcing frequency between 6 and 10 Hz. The time interval of exposure to one frequency (6 Hz) before shifting to the next frequency (10 Hz) was monotonically decreased. It was observed that the brainwave entrainment successfully shifts from one frequency value (6 Hz) to the next frequency value (10 Hz) and vice versa with the duration of exposure to each frequency being as low as 1 s. A further decrement in frequency exposure time below 1 s was not possible because of the resolution limitations posed by the sampling frequency of the instrument.

\begin{figure}[h!]
\includegraphics[width=9cm]{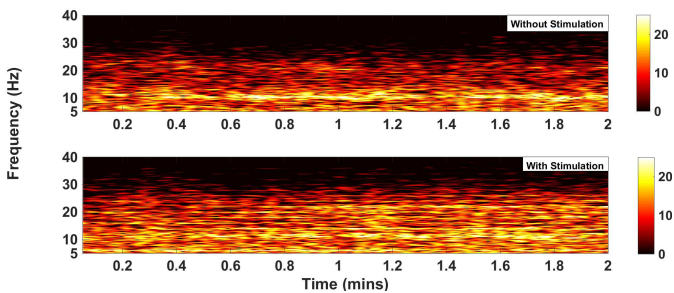}
\caption{STFT of the brainwaves (Oz electrode) for the first 2 minutes of without and with bifrequency (simultaneous 6 and 10 Hz) photic stimulation. A discernible increment in power at the fundamental frequencies (6 and 10 Hz) and the subsequent harmonics (12, 18, 20, 24, 30 and 36 Hz) can be observed from the sub-plot with the stimulus (lower panel). Also entrainment at the sum of the fundamental frequencies (6 + 10 = 16 Hz) is observed.}
\label{5}
\end{figure}

\begin{figure}[h!]
\includegraphics[width=9cm]{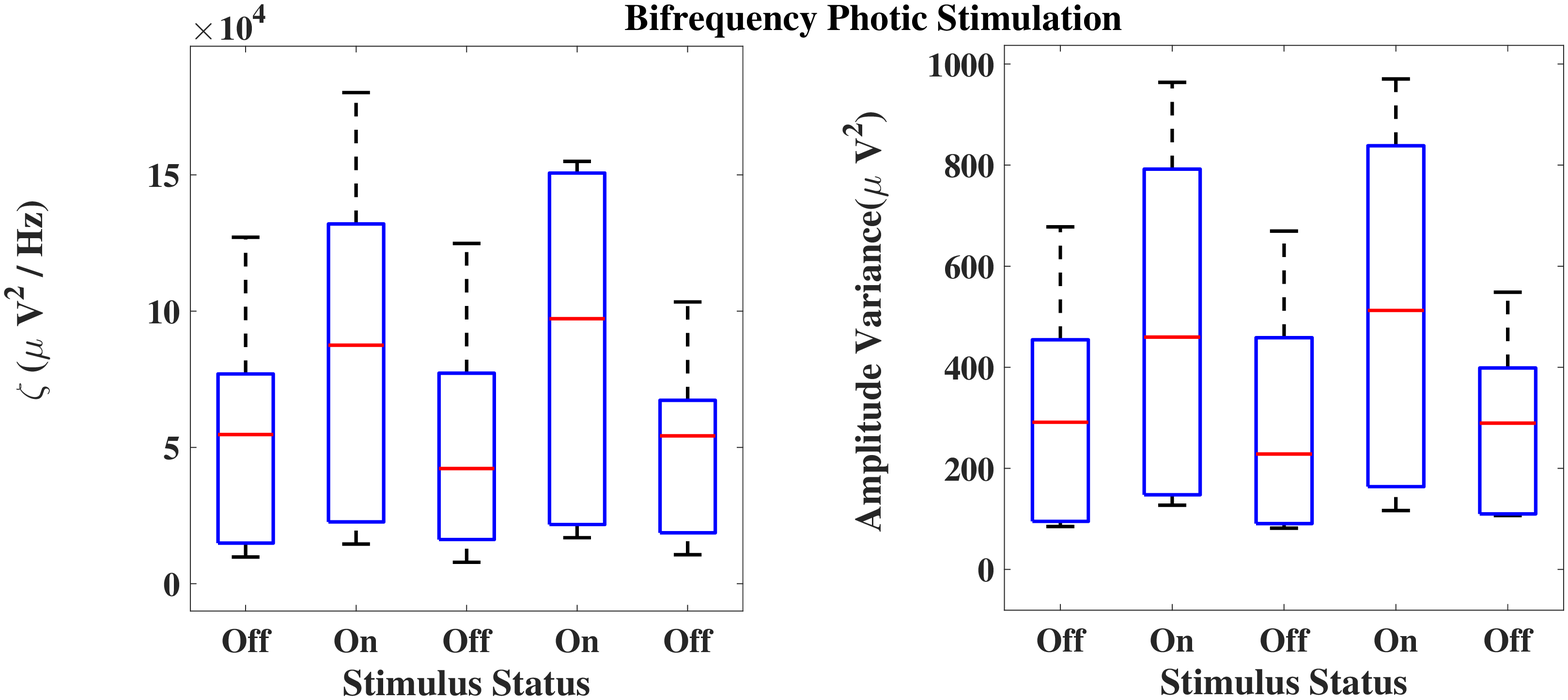}
\caption{This figure shows the variations in $\zeta$ and Amplitude Variance as the bifrequency (simultaneous 6 and 10 Hz) photic stimulation was used for entrainment. A consistent increment for both these quantities when the stimulus is provided (Stimulus Status: $\bf{On}$) can be seen from the box plots in both the figures.}
\label{6}
\end{figure}

\section{Aperiodic Entrainment}
As mentioned in the previous section, entrainment to a single frequency was possible by an exposure to that frequency for one second. Also, the entrainment shifts to the next frequency as the stimulus updates to the subsequent frequency. Hence, an aperiodic signal was made with a uniform random distribution of frequencies in the range of 5-15 Hz. The frequency of the signal kept changing randomly from one value to the next every second. This signal was fed to the LEDs and its effects on the brainwaves evaluated. As shown in Figure 7, unlike previous forms of entrainment, no consistent entrainment at a single frequency is observed. This is because the entrainment state is shifting every second due to the change in frequency values every second. In Figure 8, the increment in $\zeta$ and amplitude variance as the stimulus is provided (Stimulus Status: $\bf{On}$) persists as a consequence of entrainment. The response provoked in the brainwaves by the sequence of frequencies provided by the light signal was visually inspected. A local increment at the instantaneous stimulation frequency was observed in the brainwaves. As the stimulus changes to the next frequency, the brainwave entrainment was also found to shift towards this next frequency in the sequence. To further check if the brainwaves were following the predetermined aperiodic signal supplied through the LEDs, information transfer between the light signal and the brainwaves was quantified. In the data cleaned for this purpose, the previously rejected epochs of noisy data were replaced with zeros. This was done to maintain a uniform length of the time series. $P_k(f_i,t_i)$ represents the PSD in the frequency range of $f_i$ to $f_i + 1$ Hz and a time window of $t_i$ to $t_i + 1$ s. The time windows in this case are non-overlapping. The subscript $k$ indicates to the five parts of the experiment as mentioned in the protocol. To reiterate, during part 2 and 4, the photic stimulation was provided and part 1, 3 and 5 are without stimulation. $P_s(f_i,t_i)$ denotes the PSD for the light signal. PSD from part 1, 3 and 5 was used for defining a threshold $(th)$ which is employed as an indicator of power in a frequency band without the stimulus. The PSD was then modified as follows:\\
\[
P_k'(f_i,t_i) = \begin{cases} P_k(f_i,t_i) \hspace{1cm} \forall \hspace{0.5cm} P_k(f_i,t_i) > th \\ 0 \hspace{2.2cm} \forall \hspace{0.5cm} P_k(f_i,t_i) < th \end{cases} 
\]
$C_k$ is a measure of correlation between the STFT of the brainwaves and the light signal. It is defined as follows:\\
\[
C_k = \frac{\sum_{f_i = 5}^{15}\sum_{t_i = 0}^{400}P_k'(f_i,t_i)P_s(f_i,t_i)}{N_k}
\]

$P_s(f_i,t_i)$ and $P_k'(f_i,t_i)$ have been defined previously. First 400 s of the data was used for calculating the information transfer in the fundamental frequency range (5-15 Hz). For this analysis, $f_i$ and $t_i$ increase in steps of 1 Hz and 1 s respectively. The normalizing constant $N_k$ denotes the number of non zero data points in the brain data for the corresponding part of the experiment.

In Figure 9, an increment/decrement in $C_k$ as a function of stimulus status ($\bf{On}$/$\bf{Off}$) can be observed. To confirm $C_k$ as a reliable measure of correlation between the given aperiodic signal and the brainwaves, a comparison of the value of $C_k$ obtained using the original aperiodic signal to that of the surrogate signals was performed in Figure 10. The blue plot shows the correlation measure $C_k$ between the original aperiodic signal and the brainwaves. The other five plots measure $C_k$ between the surrogate aperiodic signals and the brainwaves. The value of $C_k$ was found to be considerably higher for the original signal when the stimulus is provided (Stimulus Status: $\bf{On}$).\\

\begin{figure}[h!]
\includegraphics[width=9cm]{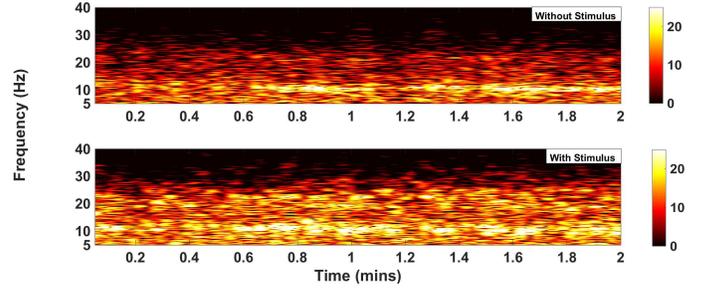}
\caption{STFT of the brainwaves (Oz electrode) for the first 2 minutes of without and with predetermined aperiodic photic stimulation. An increment in power at the fundamental frequency band (5 - 15 Hz) and the harmonic band (10 - 40 Hz) can be observed from the sub-plot with the stimulus (lower panel).}
\label{7}
\end{figure}

\begin{figure}[h!]
\includegraphics[width=9cm]{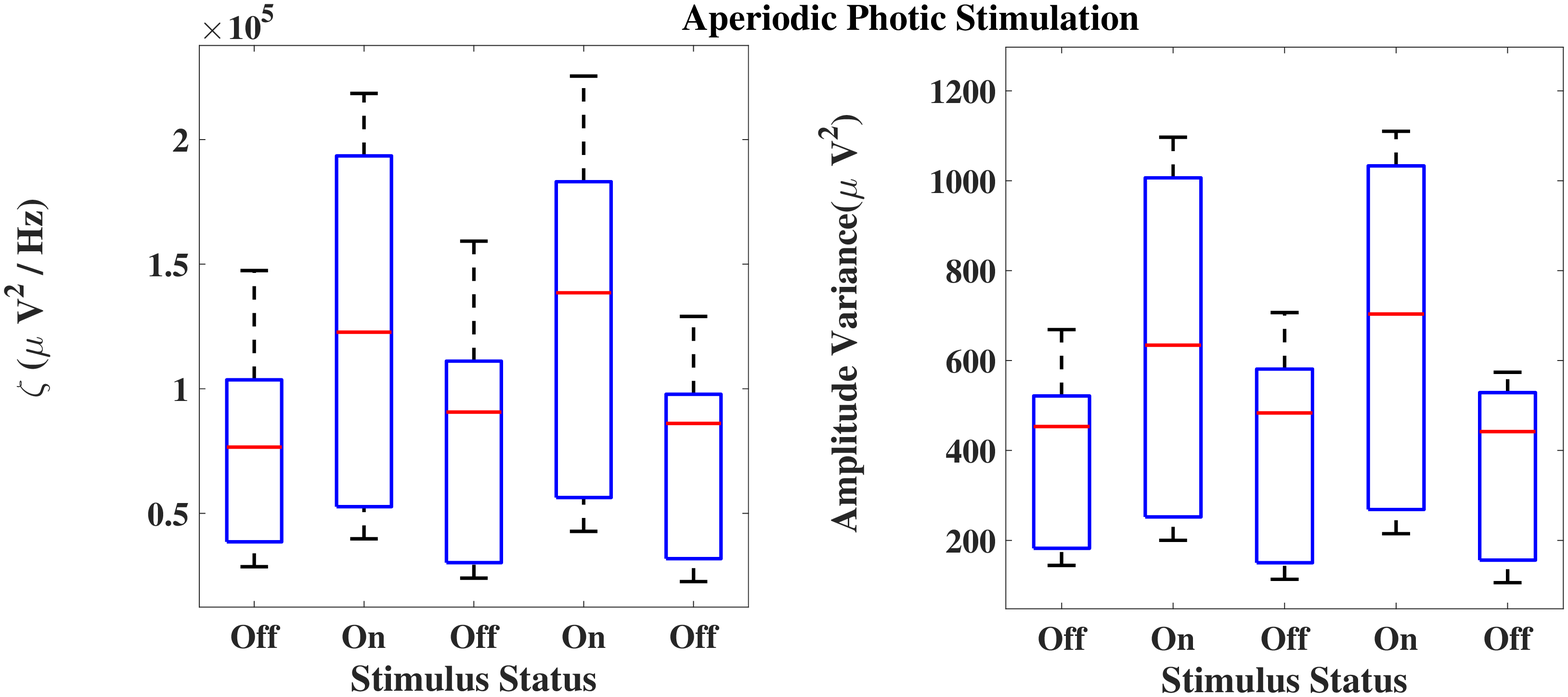}
\caption{Evolution of the quantities $\zeta$ and Amplitude Variance as a function of stimulus status for a predetermined aperiodic photic stimulus across all the subjects. A consistent increment for both these quantities when the stimulus is provided (Stimulus Status: $\bf{On}$) is evident from the box plots in both the figures.}
\label{8}
\end{figure}

\begin{figure}[h!]
\includegraphics[width=9cm]{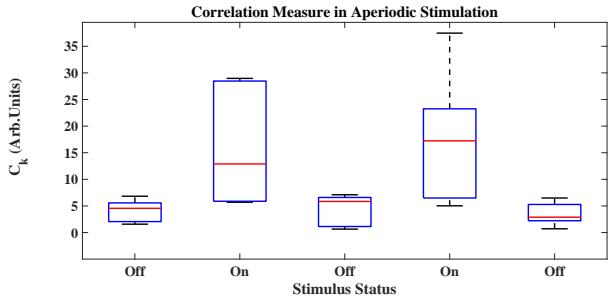}
\caption{Correlation measure ($C_k$) as a function of stimulus status. An increment in $C_k$ when the stimulus is provided (Stimulus Status: $\bf{On}$) can be seen from the figure. $C_k$ was found to be robust against the surrogate signals.}
\label{9}
\end{figure}

\begin{figure}[h!]
\includegraphics[width=9cm]{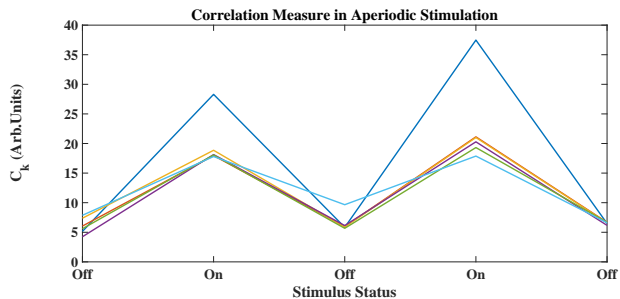}
\caption{Comparison of correlation measure ($C_k$) obtained with the surrogate signals to that obtained with the original signal. The blue plot shows $C_k$ values for the original signal. The other five plots are the $C_k$ values for the surrogate data. It can be seen from the graph that $C_k$ values for the original signal when the stimulus is provided (Stimulus Status: $\bf{On}$) are considerably higher than those for the surrogate data.}
\label{10}
\end{figure}

\section{Discussion}
In the results presented above, entrainment is found to persist with different types of stimulus including periodic, bifrequency and aperiodic photic stimulation. The STFT plots across various forms of stimulus show entrainment for one subject while the box plots represent the robustness of the phenomenon against subject variability. The decrement in $\zeta$ and amplitude variance after the stimulus is removed is a manifestation of the underlying information processing area. The effect of stimulation dies down after the system stops receiving the external information. The increment in amplitude variance of the EEG signal when the stimulus is provided is an indicator of increased neuronal firing in the corresponding region of the brain. This can be loosely compared to a stimulus provoked increased blood flow in a specific brain region observed using various neuroimaging techniques. In periodic entrainment, as shown using the power spectral density (PSD) in the supplementary material (Supplementary Figure 5 and 6), the increment in power is relatively higher with the forcing frequency of 10 Hz as compared to the 6 Hz stimulation. This is in agreement with~\cite{Notbohm}, where maximum strength of entrainment for a specified light intensity is observed when the frequency mismatch between the dominant alpha frequency and the forcing frequency is minimum. Therefore, the absence of sub-harmonic entrainment in case of 6 Hz stimulation can be explained by the relative weakness of entrainment observed with 6 Hz forcing frequency. In bifrequency entrainment, a simultaneous increment in power at the fundamental, harmonic and summation of frequencies was observed. Aperiodic entrainment, in our opinion, is of special interest as it might have wide applicability. One possible application could be in the field of biofeedback. The brainwaves of a healthy individual during various stages could be recorded and fed back to them as a visual stimulus. In certain pathological conditions such as some forms of epilepsy (Petit Mal Seizures), low frequency rhythmic activity in the brainwaves is observed. An aperiodic pattern lying outside this rhythmic activity and centred near the baseline state of the subject might lead to the modification of the brainwaves to the desired state. An aperiodic entrainment of circadian rhythms may also be looked into. 

\section{Supplementary Material}
In the supplementary material, the results of the sub-harmonic entrainment for the two periodic photic stimulations are reported. A detailed comparison between the periodic auditory and periodic visual entrainment using white sound and white light stimulus respectively is also presented in the supplementary material.

\section{Acknowledgements}
The authors would like to thank DST (India), (Project ref no. EMR/2016/000275) for financial assistance. Richa Phogat would also like to acknowledge CSIR (India) for financial assistance. We would also like to acknowledge Guru Vamsi Policharla, Kanishk Chauhan, Keshav Srinivasan and all the past and present members of the NLD lab group of IIT-Bombay for help during various stages of the study.

\end{document}


\makeatletter
\title{Supplementary Material}
\maketitle

\section{Sub-harmonic entrainment for periodic photic stimulus}
The sub-harmonic entrainment for periodic photic stimulation was observed only for 10 Hz and not for 6 Hz stimulation. This is shown in Figure 1 where the data for Oz electrode is filtered in the frequency range of 4.9 - 5.1 Hz to study the sub-harmonic entrainment of 10 Hz photic stimulation (5 Hz). As shown by the consistent increment in $\zeta$ and amplitude variance when the stimulus is provided (Stimulus Status: $\bf{on}$ in Figure 1), the sub-harmonic entrainment for 10 Hz is observed across all subjects. For 6 Hz photic stimulation, the data was filtered in the range of 2.9 - 3.1 Hz (sub-harmonic for 6 Hz at 3 Hz). As shown in Figure 2, this pattern of increment in $\zeta$ and amplitude variance is not universally observed across all the subjects.

\begin{figure}
\includegraphics[width=9cm]{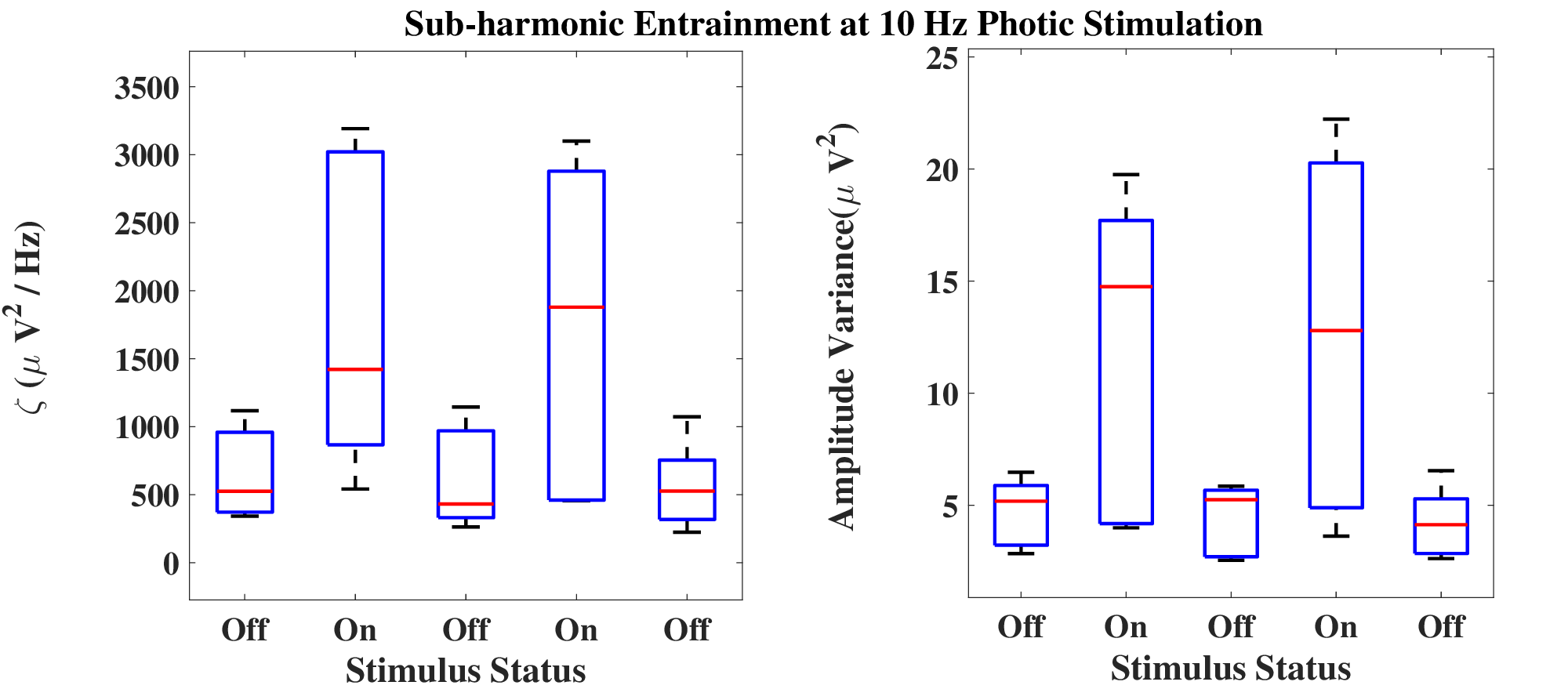}
\caption{Evolution of the quantities $\zeta$ and amplitude variance at sub-harmonic frequency as a function of stimulus status for all the subjects for 10 Hz photic stimulation. A universal increment for both these quantities when the stimulus is provided (Stimulus Status: $\bf{on}$) is evident from the box plots in both the sub-figures.}
\label{1}
\end{figure}

\begin{figure}
\includegraphics[width=9cm]{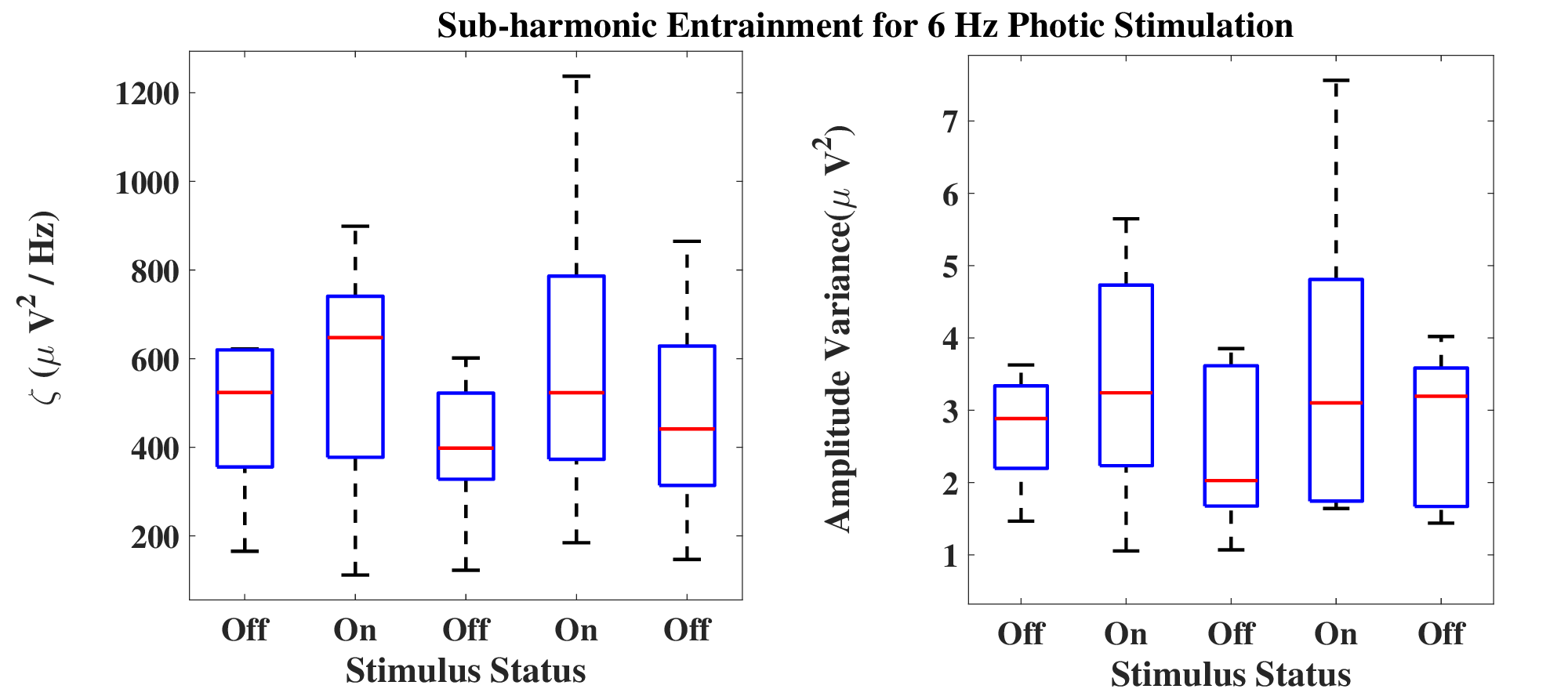}
\caption{Evolution of the quantities $\zeta$ and amplitude variance at sub-harmonic frequency as a function of stimulus status for all the subjects for 6 Hz photic stimulation. A universal increment in both these quantities when the stimulus is provided (Stimulus Status: $\bf{on}$) is lacking as can be seen from the box plots in both the sub-figures.}
\label{2}
\end{figure}

\section{Periodic auditory stimulation}
To study auditory entrainment, the evolution of the electrical activity of brain as captured by electrodes T5 and T6 in the temporal head region was followed. A case by case comparison of the auditory entrainment with the photic entrainment is presented here to give a better idea of the entrainment. Figure 3 and 4 show the Power Spectral density (PSD) for the auditory entrainment with a white sound. The stimulation provided consists of a square signal oscillating at a desired frequency between white sound and no sound states. For comparison, the PSD for photic entrainment with a white light flickering at 10 and 6 Hz respectively is also presented in Figure 5 and 6 for the same subject. The photic entrainment is studied in the occipital head region i.e. for the O1, O2 and Oz electrodes. 

\begin{figure}
\includegraphics[width=9cm]{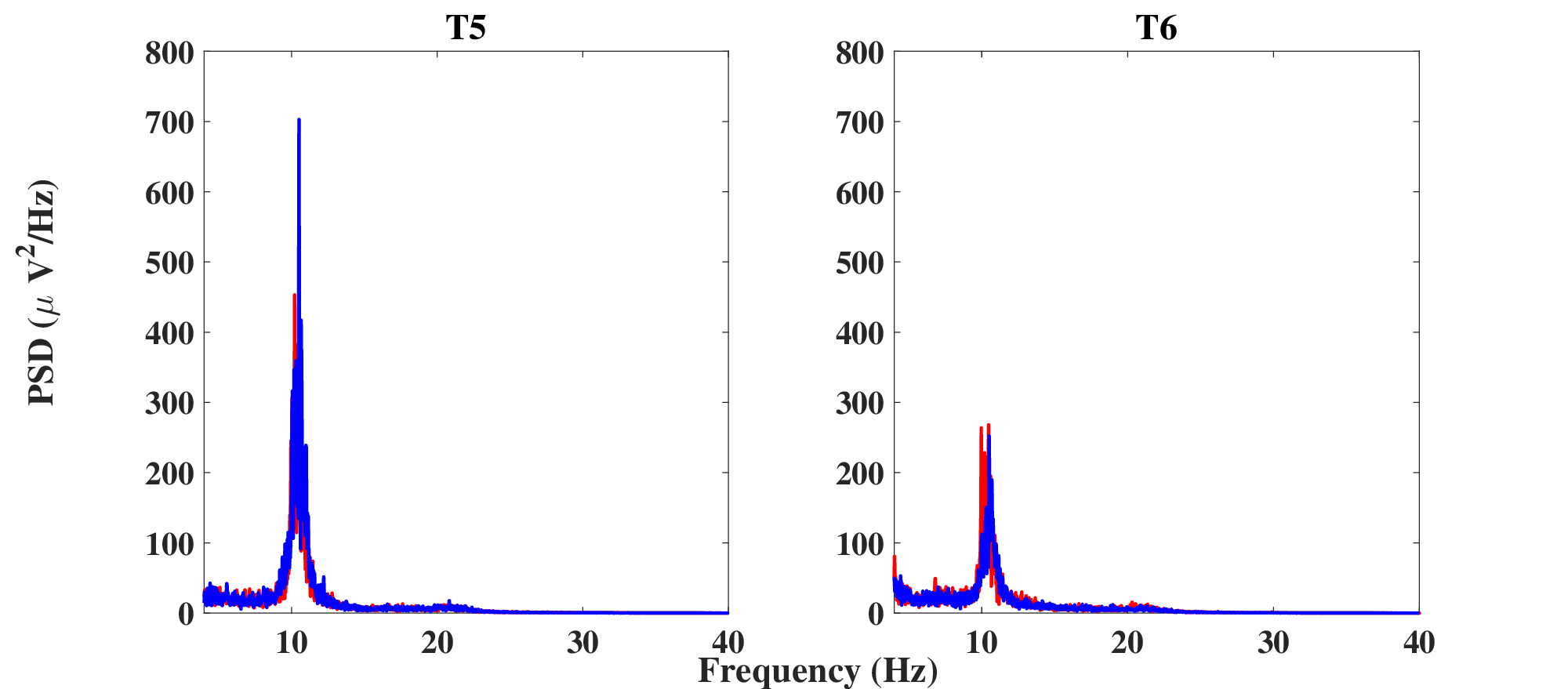}
\caption{PSD as a function of frequency for the two electrodes T5 and T6. The stimulation in this case was given in the form of white sound. The red and blue plots represent the data with and without stimulation respectively. The without stimulation data belongs to part III where the stimulation was off after stimulation being provided during part II of the experimental protocol. Entrainment was observed only for the fundamental frequency of 10 Hz in T6 electrode. No discernible change in the power at T5 electrode was observed.}
\label{3}
\end{figure}

\begin{figure}
\includegraphics[width=9cm]{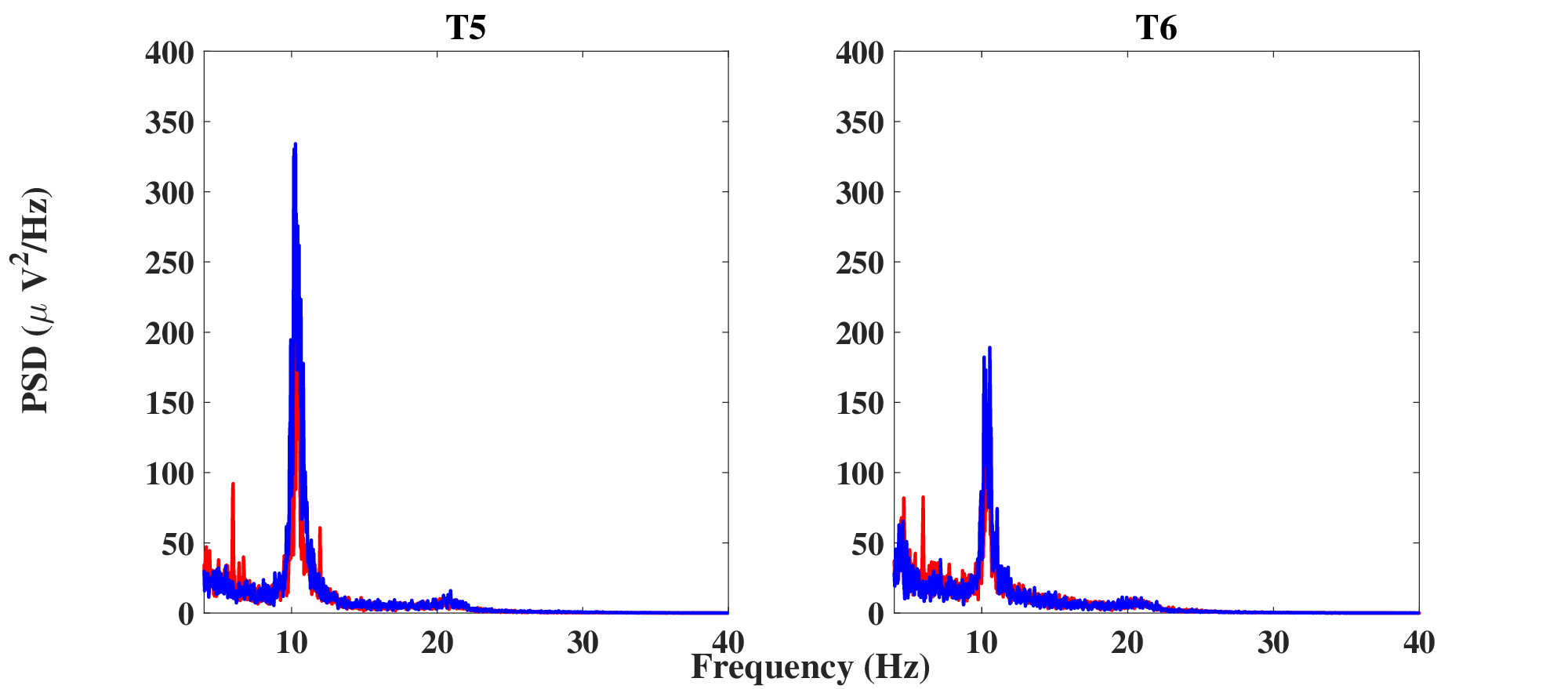}
\caption{PSD as a function of frequency for the two electrodes T5 and T6. The stimulation in this case was given in the form of white sound oscillating at 6 Hz frequency. The red and blue plots represent the data with and without stimulation respectively. The without stimulation data belongs to part III where the stimulation was off after the stimulation being provided during part II. Entrainment was observed for the fundamental frequency and second harmonic (6 and 12 Hz) in T5 electrode. For T6 electrode, only fundamental frequency entrainment was observed.}
\label{4}
\end{figure}

\begin{figure}
\includegraphics[width=9cm]{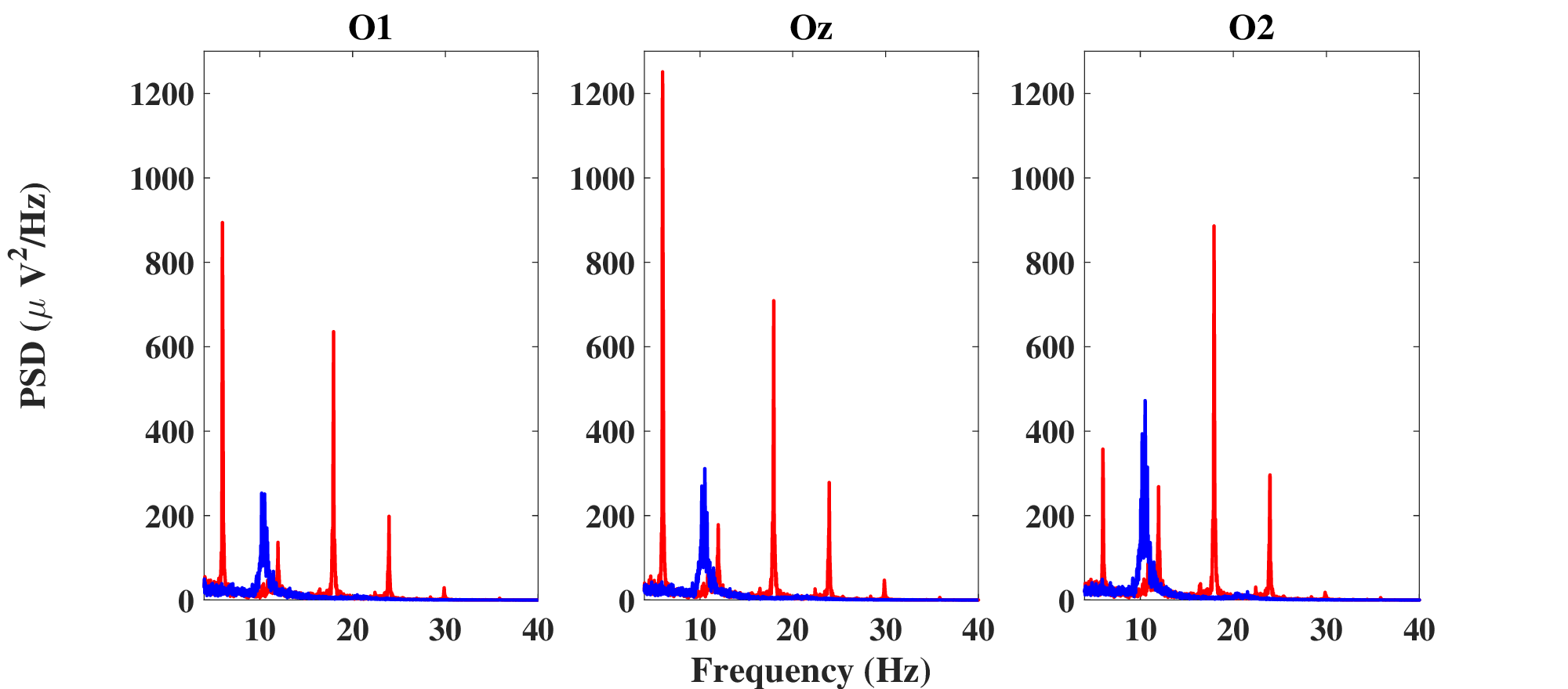}
\caption{PSD as a function of frequency for the three electrodes O1, O2 and Oz. A peak at 6,12,18,24,30 and 36 Hz shows entrainment to the 6 Hz photic stimulation. This stimulation was given in the form of white light. The red and blue plots represent the data with and without stimulation respectively. The without stimulation data belongs to part III where the stimulation was off after an exposure to stimulation during part II. The absence of this peak in the blue curve indicates to the attenuation of the entrainment after the stimulus is removed.}
\label{5}
\end{figure}

\begin{figure}
\includegraphics[width=9cm]{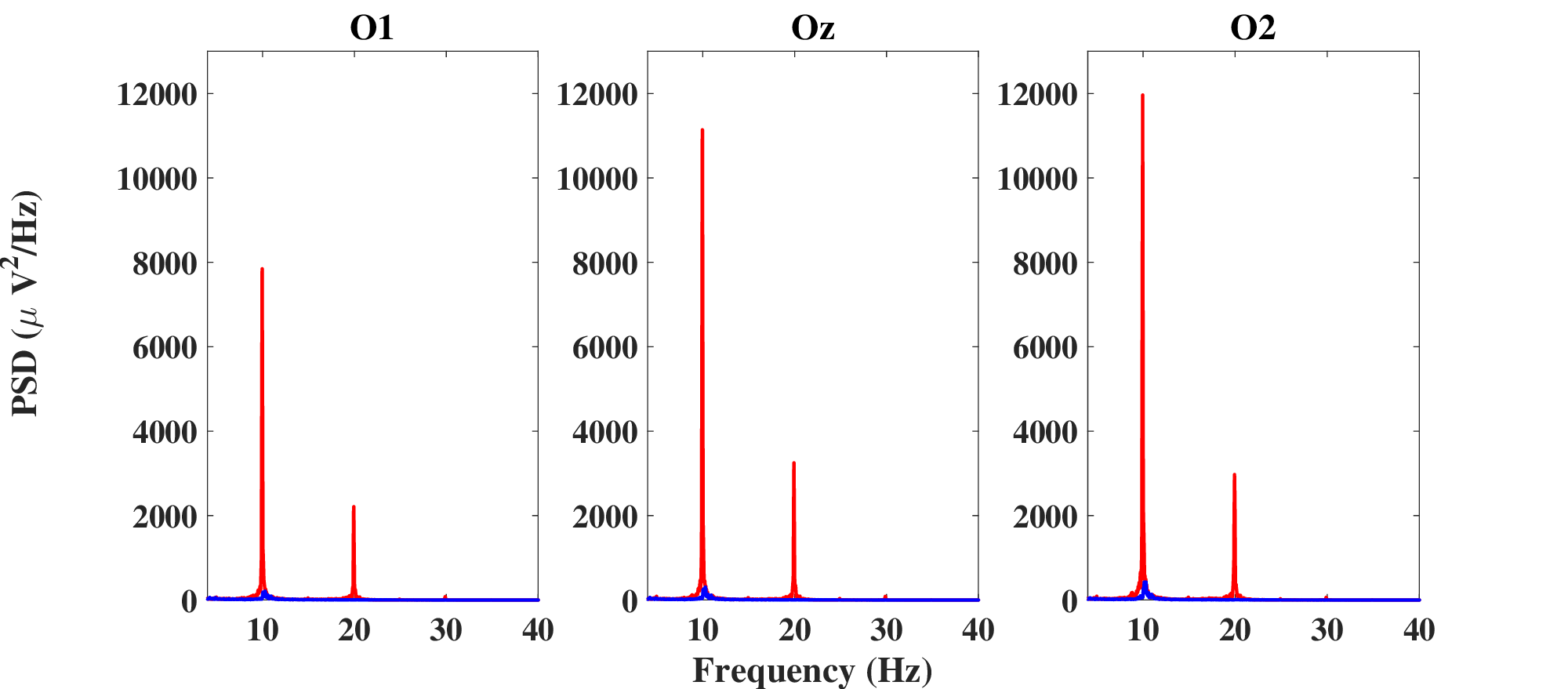}
\caption{PSD as a function of frequency for the three electrodes O1, O2 and Oz. A peak at 10,20 and 30 Hz shows entrainment to the 10 Hz photic stimulation. This stimulation was given in the form of white light. The red and blue plots represent the data with and without stimulation respectively. The without stimulation data belongs to part III where the stimulation was off an exposure during part II. The absence of this peak in the blue curve indicates to the attenuation of the entrainment after the stimulus is removed.}
\label{6}
\end{figure}

Upon comparing the power spectrum for auditory and photic stimulation, it is indicated that photic stimulus provokes superior entrainment as compared to its auditory counterpart. In addition, entrainment by auditory stimulation was not observed across all the subjects.